\documentclass{iau}

\usepackage{amsmath}
\usepackage{graphicx}
\usepackage{multirow}
\usepackage{array}
\usepackage{url}
\usepackage{placeins}
\usepackage{ragged2e}

\makeatletter
\newcommand{\NAT@bib@parstart}{%
  \ifhmode\unskip\par\fi
  \addvspace{\bibsep}%
  \RaggedRight
  \noindent\hangindent\bibhang\hangafter=1\relax
}%
\renewenvironment{thebibliography}[1]{%
  \bibsection
  \parindent\z@
  \bibfont
  \global\c@NAT@ctr\z@%
  \fussy\clubpenalty10000\widowpenalty10000
  \sfcode`\.\@m
  \let\NAT@bibitem@first@sw\@firstoftwo
    
 }{%
  \bibitem@fin
  \bibpostamble
  \def\@noitemerr{%
    \PackageWarning{natbib}{Empty `thebibliography' environment}}%
  \bibcleanup
}%
\def\@lbibitem[#1]#2{%
  \if\relax\@extra@b@citeb\relax\else
    \@ifundefined{br@#2\@extra@b@citeb}{}{%
     \@namedef{br@#2}{\@nameuse{br@#2\@extra@b@citeb}}%
    }%
  \fi
  \@ifundefined{b@#2\@extra@b@citeb}{%
   \def\NAT@num{}%
  }{%
   \NAT@parse{#2}%
  }%
  \def\NAT@tmp{#1}%
  \expandafter\let\expandafter\bibitemOpen\csname NAT@b@open@#2\endcsname
  \expandafter\let\expandafter\bibitemShut\csname NAT@b@shut@#2\endcsname
  \@ifnum{\NAT@merge>\@ne}{%
   \NAT@bibitem@first@sw{%
    \@firstoftwo
   }{%
    \@ifundefined{NAT@b*@#2}{%
     \@firstoftwo
    }{%
     \expandafter\def\expandafter\NAT@num\expandafter{\the\c@NAT@ctr}%
     \@secondoftwo
    }%
   }%
  }{%
   \@firstoftwo
  }%
  {%
   \global\advance\c@NAT@ctr\@ne
   \@ifx{\NAT@tmp\@empty}{\@firstoftwo}{%
    \@secondoftwo
   }%
   {%
    \expandafter\def\expandafter\NAT@num\expandafter{\the\c@NAT@ctr}%
    \global\NAT@stdbsttrue
   }{}%
   \bibitem@fin
   \NAT@bib@parstart
   \NAT@anchor{#2}{}%
   \global\let\NAT@bibitem@first@sw\@secondoftwo
   \NAT@bibitem@init
  }%
  {%
   \RaggedRight
   \NAT@anchor{#2}{}%
   \NAT@bibitem@cont
   \bibitem@fin
  }%
  \@ifx{\NAT@tmp\@empty}{%
    \NAT@wrout{\the\c@NAT@ctr}{}{}{}{#2}%
  }{%
    \expandafter\NAT@ifcmd\NAT@tmp(@)(@)\@nil{#2}%
  }%
}%
\makeatother

\begin{document}

\lefttitle{Sanjoy M. Som}
\righttitle{The spacefaring envelope}

\jnlPage{1}{6}
\jnlDoiYr{2024}
\doival{10.1017/xxxxx}

\aopheadtitle{Proceedings IAU Symposium No.~404, 2024}
\editors{J. Haqq-Misra \& R. Kopparapu, eds.}

\title{Multistage Rocket Optimization, Geophysics, and the
Spacefaring Envelope of Habitable Super-Earths}

\author{Sanjoy M. Som}
\affiliation{Blue Marble Space \\ email: {\tt sanjoy@bmsis.org}}

\begin{abstract}
Habitability studies traditionally ask whether a planet can host biology; they
rarely ask whether it can host a civilization able to leave it. In this work,
we treat spacefaring capability as a distinct, technological axis of habitability,
defined operationally as placing a 1000~kg payload on an escape
trajectory using chemical propulsion, a deliberately
conservative, ``Voyager-class'' benchmark. We develop a coupled
geophysical--atmospheric--astronautical framework that maps a ``spacefaring envelope'' onto planetary mass and surface air pressure axes.  Building on
\citet{Hippke2018} and \citet{Gonzalez2020}, we optimize a multistage chemical rocket whose stage count minimizes the reliability-weighted expected launch mass, and we report first-stage engine number, stage count, and mission reliability. We validate against flown vehicles, assuming propulsion is driven by an engine akin to the Saturn V F-1 first-stage engine. Despite simplifying assumptions, Saturn~V gross lift-off mass matches the model to $\sim\!30\%$ and the F-1 turbopump power to within $\sim\!18\%$. Over $0.1$--$10$~bar, surface pressure changes launch mass by up to $\sim\!35\%$ on $0.5\,M_\oplus$ planets where drag is a larger fraction of the ascent $\Delta v$ budget, but by only a few percent for $M_\oplus \gtrsim 4$. Within our assumptions, gravity, not atmospheric drag, limits rocket escape from super-Earths. A first-stage clustering limit of $\sim\!100$ F-1-class engines imposed \emph{after} optimization renders chemical escape of the benchmark payload impractical above $\sim\!11.5\,M_\oplus$. This explicit
engine-counting argument independently corroborates the $\sim\!10\,M_\oplus$
ceiling that \citet{Hippke2018} derived from an engine-independent fuel-ratio
argument. These results provide a physically motivated metric for
exoplanets where a technological civilization could plausibly escape its gravity
well.
\end{abstract}


\maketitle

\section{Introduction}

Habitability studies have traditionally emphasized conditions that support
biology: liquid water, raw materials (CHNOPS), energy, and biophysically plausible environmental conditions \citep{Hoehler2007}. Whether a world
can also host a civilization able to explore its planetary system is a distinct
and largely unexplored question. We refer to this extension as
\textit{spacefaring habitability,} in contrast to \textit{technological habitability} defined by \citet{cockell2016} as a civilization's ability to modify its planetary environment to ensure survival. Deep-space missions have been linked to a broader shift in planetary awareness \citep{White1987,Yaden2016} and may contribute to a civilization's urge to keep exploring and expanding its spacefaring capabilities (an admittedly anthropocentric perspective). Here, we focus on the physical requirements for spacefaring capability.

For Earth, this threshold was arguably crossed with spacecraft of order
$\sim10^3$~kg, most prominently the Voyager probes. Voyager opened humanity's eyes to the diversity of possible worlds and pushed forward the question ``Are we alone?'' Although Voyager's mass reflects contingent engineering, the $\sim10^3$~kg scale is not arbitrary: smaller payloads struggle to carry the communications, power, propulsion, and instrument suites needed for robust comparative planetology, whereas larger payloads incur severe energetic penalties because launch mass scales exponentially with $\Delta v$ (as derived from the rocket, or Tsiolkovsky, equation). Throughout, ``Voyager-class'' denotes a 1000~kg payload on a planetary escape trajectory.

The physical difficulty of leaving a more massive world has previously been
explored. \citet{Hippke2018} showed, using the rocket equation, that the fuel-to-payload ratio scales exponentially with surface gravity for chemical rockets, and concluded that chemical escape remains possible up to approximately $10\,M_\oplus$.
\citet{Gonzalez2020} extended the discussion to atmospheric pressure, re-entry,
and the broader Solar System context. Stellar escape, in contrast to planetary escape, from the habitable zones of low-mass stars can impose additional constraints \citep{Loeb2018,LingamLoeb2018}. In this contribution, we deliberately treat the free-flying vehicle by itself in order to obtain a conservative lower bound on technological capability and do not explore other means of launch, such as magnetic levitation rails for propelling payloads to escape velocity \citep{powell2010}.

Despite extensive work on exoplanet habitability and planetary geodynamics
\citep[e.g.,][]{Foley2015,Valencia2006}, we are aware of no framework that
jointly optimizes multistage launch vehicles against planetary interior, planetary mass, and surface air pressure. Addressing that gap requires coupling interior and
atmospheric scaling with chemical-rocket energetics, in which required wet mass
scales exponentially with total $\Delta v$ (the sum of escape speed, gravity losses, and
drag losses). Here, we develop such a coupled geophysical--atmospheric--energetic framework. Using a simplified framework, we assess the plausibility of plate tectonics and the dynamo from the mantle Rayleigh and magnetic Reynolds numbers \citep{Valencia2006,Valencia2007,Christensen2010,Foley2015}. We
incorporate drag through an exponentially decreasing 80\%~N$_2$/20\%~O$_2$ atmosphere and
size multistage rockets for which
$\Delta v = v_{\rm esc}+\Delta v_{\rm grav}+\Delta v_{\rm drag}$. Because any
$\Delta v$ is reachable in principle with enough stages, feasibility is assessed
with engineering constraints: we select the stage count that minimizes the
reliability-weighted expected launch mass (Sec.~\ref{sec:reliability}), then
apply hardware constraints on engine clustering and total launch mass
(Sec.~\ref{sec:assumptions}). Assumptions are collected in
Table~\ref{tab:assumptions}.

\section{Methods}
\label{sec:methods}

Our framework couples planetary geophysics, atmospheric structure, and chemical-rocket astronautics. Simple geophysical scalings label plate tectonics and a core dynamo as robustt, weak, or unlikely. These scalings should not be taken as authoritative; rather, they provide a first pass tuned to Earth, and they are not a substitute for physically consistent numerical convection models. The astronautical model is the quantitative core of this work. We test the optimized vehicles against engineering limits tied to maximum rocket mass and engine clustering that we expect to be approximately universal for chemical propulsion. Indeed, it is these limits that define the \emph{spacefaring envelope} across the parameter space of planetary mass and surface air pressure.

\subsection{Model assumptions and operational definitions}
\label{sec:assumptions}

Table~\ref{tab:assumptions} lists the assumptions used in the numerical model
\citep{spacefaring_code}. The payload is fixed at $m_{\rm pay} = 1000$~kg on a planetary escape trajectory (Sec.~\ref{sec:results}); this is our operational Voyager-class benchmark. The gross lift-off mass (GLOM) $m_0$ is the sum of all stage wet masses plus the payload. Staging is optimized to minimize the expected
launch mass per successful mission (Sec.~\ref{sec:reliability}). The engine-count and
``Cheops pyramid'' mass limit of \citet{Hippke2018} are applied \emph{after} the optimization.

\begin{table}[t]
\centering
\caption{Modeling assumptions and their role in the astronautical model.
``Optimization'' entries set the vehicle design; Constraints test
feasibility of the optimized design.}
\label{tab:assumptions}
{\tablefont\small
\begin{tabular}{@{}p{0.34\linewidth}p{0.58\linewidth}@{}}
\midrule
Assumption & Role / value \\
\midrule
Payload & $m_{\rm pay}=1000$~kg, escape trajectory \\
Mass--radius & $R=R_\oplus(M/M_\oplus)^{0.27}$ [Eq.~(\ref{eq:radius})] \\
Atmosphere & 80\%~N$_2$/20\%~O$_2$; $T_{\rm surf}=288$~K; $P_{\rm surf}$ variable \\
Rotation & No $\Delta v$ credit from planetary spin (conservative) \\
Ascent trajectory & Vertical; constant net acceleration $a_{\rm net}=0.5\,g$ (TWR $=1.5$) \\
Drag area $S$ & Cylindrical first-stage tank; AR $=3$; engines omitted from $S$ [Eqs.~(\ref{eq:drag})--(\ref{eq:expatmo})] \\
Drag mass & Mean vehicle mass $\langle m\rangle = m_0/2$ during ascent \\
$I_{\rm sp}$ staging & Top two stages 450~s (LH$_2$/LOX); lower stages 350~s (RP-1/LOX) \\
Structure & $\epsilon=0.10$ for all stages [Eq.~(\ref{eq:mwet})] \\
$\Delta v$ split & Equal per stage \\
Gravity loss & $\Delta v_{\rm grav}=1500\,(g/g_\oplus)^{1/2}$~m\,s$^{-1}$ [Eq.~(\ref{eq:dvbudget})] \\
Drag loss & Integrated; seeded at $118\,P_{\rm surf}$~m\,s$^{-1}$, iterated with $m_0$ \\
Stage reliability & $R_s=0.97$ per stage (optimization) \\
Engine reference & F-1 class: $\dot m_{\rm F\text{-}1}=2600$~kg\,s$^{-1}$, pump $\Delta P_0\sim 100$~bar \\
Engine $I_{\rm sp}$ at liftoff & 350~s (RP-1/LOX lower-stage value; see text) \\
Engine count & $N_{\rm eng}=\lceil P_{\rm turbo,tot}/P_{\rm turbo,eng}\rceil$ [Eq.~(\ref{eq:turbo})] \\
Constraint: clustering & $N_{\rm eng}\le 100$ (first stage) \\
Constraint: launch mass & $m_0\le 4\times 10^5$~t; ``Cheops" launch mass limit \citep{Hippke2018} \\
Geophysics & Non-dimensional treatment (Sec.~\ref{sec:geophys}) \\
\midrule
\end{tabular}}
\end{table}

\subsection{Geophysical screen}
\label{sec:geophys}
We assess plate-tectonic and dynamo plausibility from three criteria: the mantle
Rayleigh number $Ra$ (vigor of mantle convection), the core magnetic Reynolds
number $Rm$ (dynamo activity), and the planetary mass itself. Both very small,
$<0.5\,M_\oplus$, and very large, $>10\,M_\oplus$, worlds may be disfavored due to different convection regimes \citep{Valencia2007}. Planetary radius follows the terrestrial mass--radius relation of \citet{Valencia2006}, where the subscript $\oplus$ indicates Earth,
\begin{equation}
R = R_\oplus (M/M_\oplus)^{\beta}, \qquad \beta = 0.27.
\label{eq:radius}
\end{equation}
This relationship applies to planets with Earth-like compositions and accounts for the compressibility of more massive planets.

\subsubsection{Mantle convection}
\begin{figure}[t]
\centering
\includegraphics[width=0.82\linewidth]{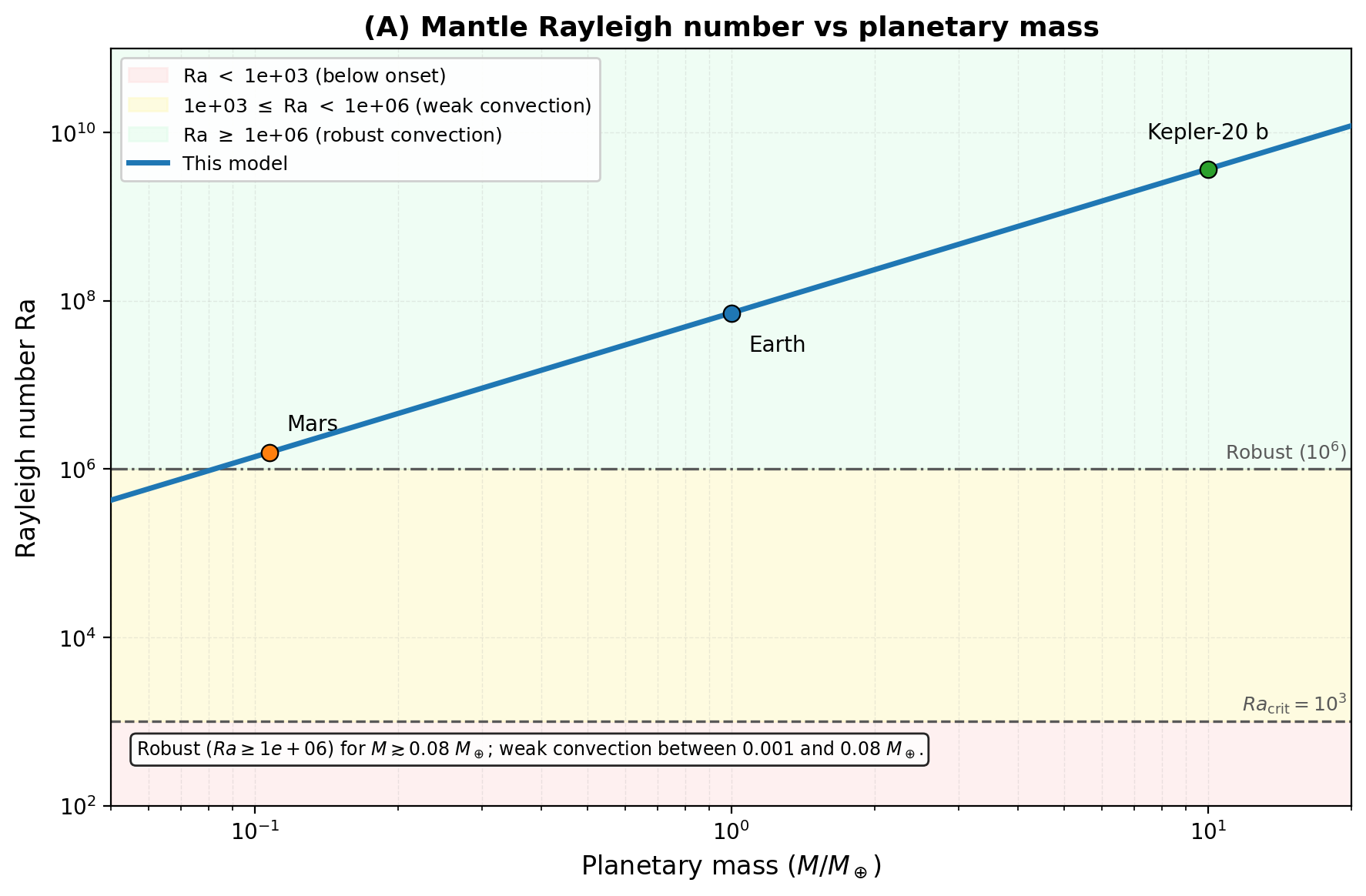}\\[0.6em]
\includegraphics[width=0.82\linewidth]{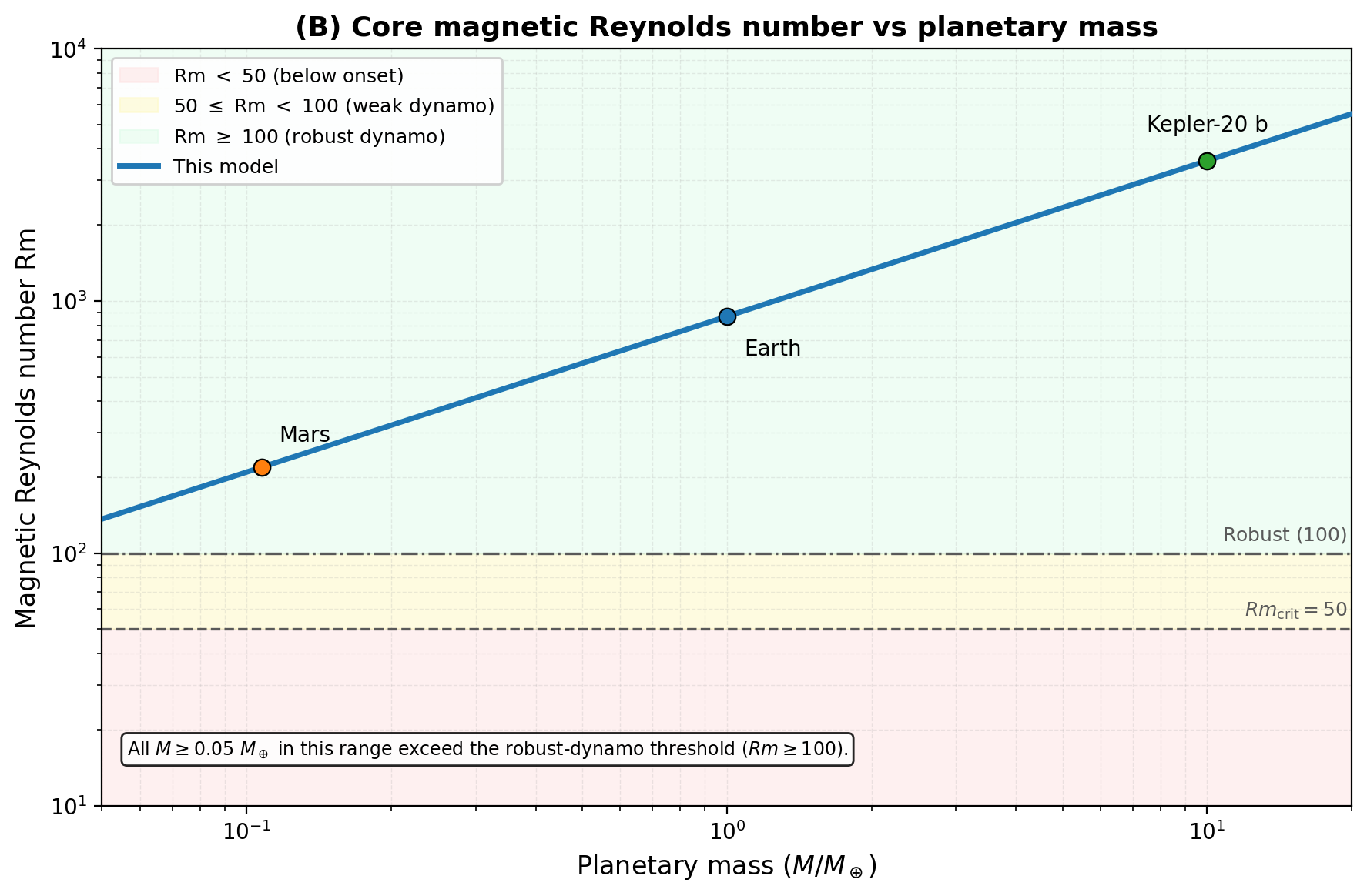}
\caption{Geophysical constraints from Sec.~\ref{sec:geophys} as a function
of planetary mass (log--log axes). \textbf{(A)}~Mantle Rayleigh number $Ra$ with
onset ($Ra=10^{3}$) and robust-convection ($Ra=10^{6}$) bands.
\textbf{(B)}~Core magnetic Reynolds number $Rm$ with dynamo-onset
($Rm=50$) and robust-dynamo ($Rm=100$) bands. Blue curves: this model; markers:
Earth, Mars ($M/M_\oplus \approx 0.11$), and Kepler-20~b ($10\,M_\oplus$,
$1.7\,R_\oplus$; \citealt{Hippke2018}). At Mars the model gives $Ra \approx 1.6\times10^{6}$ and
$Rm \approx 220$, i.e.\ robust convection and a robust dynamo under the adopted
scalings, even though Mars today lacks plate tectonics and an active global
magnetic field (an ancient dynamo is recorded in crustal magnetization). Mars
therefore illustrates the limit of this mass-only, equilibrium approach: $Ra$
and $Rm$ test convection vigor and dynamo capacity, not tectonic mode or dynamo
longevity. Kepler-20~b is a super-Earth ($Ra \approx 3.7\times10^{9}$, $Rm \approx 3.6\times10^{3}$) and passes both
thresholds.}
\label{fig:geophys}
\end{figure}

The Rayleigh number is
\begin{equation}
Ra = \frac{\rho\, g\, \alpha\, \Delta T\, d^{3}}{\kappa\, \eta},
\label{eq:rayleigh}
\end{equation}
with mantle density $\rho$, surface gravity $g$, thermal expansivity
$\alpha = 3\times10^{-5}$~K$^{-1}$, temperature contrast $\Delta T$, mantle
thickness $d$, thermal diffusivity $\kappa = 10^{-6}$~m$^2$\,s$^{-1}$, and
dynamic viscosity $\eta$. Assuming internal heating at the terrestrial rate,
which fixes terrestrial reference values ($\rho_\oplus = 4000$~kg\,m$^{-3}$,
$d_\oplus = 2900$~km, $\Delta T_\oplus = 2500$~K;
\citealt{TurcotteSchubert2002}), the mass-dependent quantities scale as
\citep{Valencia2007}
\begin{equation}
\rho = \rho_\oplus \mu^{0.2}, \qquad
g = g_\oplus \mu^{0.5}, \qquad
d = d_\oplus \mu^{0.28}, \qquad \mu \equiv M/M_\oplus .
\label{eq:scalings}
\end{equation}
The mantle temperature contrast follows from boundary-layer theory. We adopt the
Nusselt--Rayleigh relation $Nu \sim Ra^{\gamma}$ with $\gamma = 1/3$
\citep[][p.~273]{TurcotteSchubert2002} (the alternative $\gamma = 1/2$ of
\citealt{ChristensenAubert2006}, their Eq.~43, would change the exponents below).
With the conductive flux $q_{\rm cond} = k\,\Delta T/d \sim \Delta T/d$, the total
flux is
\begin{equation}
q_{\rm tot} = Nu\,q_{\rm cond} \sim Ra^\gamma \Delta T/d \sim (\rho g)^\gamma \Delta T^\gamma d^{3\gamma}\Delta T d^{-1}. 
\end{equation}
For $\gamma = 1/3$ the mantle-thickness powers cancel ($d^{3\gamma}d^{-1}=d^{0}$),
so $q_{\rm tot}\sim(\rho g)^{1/3}\Delta T^{4/3}$ is independent of $d$. Solving for
$\Delta T$ gives
\begin{equation}
\Delta T \sim \left(\frac{q_{tot}}{(\rho g)^{1/3}}\right)^{3/4}\sim q_{\rm tot}^{\,3/4}\,(\rho g)^{-1/4}
\label{eq:dT}
\end{equation}
The heat flux is the internal power per unit area at the core-mantle boundary. Assuming that the internal power $\sim$ planetary mass $M$, $q_{\rm tot}\sim M/R^{2}\sim M/M^{0.54}\sim M^{0.46}$ using Eq.~(\ref{eq:radius}); substituting into Eq.~(\ref{eq:dT}) with Eq.~(\ref{eq:scalings}) gives $\Delta T \sim (M^{0.46})^{3/4}(M^{0.2}M^{0.5})^{-1/4}$, or  $\Delta T = \Delta T_\oplus\,\mu^{0.17}$, independent of $d$. Combining Eqs.~(\ref{eq:scalings})--(\ref{eq:dT}) in Eq.~(\ref{eq:rayleigh}) under the isoviscous assumption ($\eta = 10^{21}$~Pa\,s, justified because the sub-lithospheric temperature is nearly mass-independent; \citealt{Valencia2007}) yields $Ra\sim M^{0.2}M^{0.5}M^{0.17}(M^{0.28})^3$, or
\begin{equation}
Ra = Ra_\oplus\,\mu^{1.71}, \qquad Ra_\oplus = 7\times10^{7} .
\label{eq:rascale}
\end{equation}
We take the classical onset of convection at $Ra_{\rm crit} = 10^{3}$ and
``robust'' convection at $Ra = 10^{6}$ (Fig.~\ref{fig:geophys}a).

\subsubsection{Core dynamo}
We use the magnetic Reynolds number as a dynamo proxy,
\begin{equation}
Rm = \frac{U_{\rm core}\,R_{\rm core}}{\eta_{\rm mag}},
\qquad R_{\rm core} = R - d,
\label{eq:rm}
\end{equation}
with magnetic diffusivity $\eta_{\rm mag} = 2.0$~m$^2$\,s$^{-1}$ and a critical
value $Rm_{\rm crit} = 50$ for an active dynamo. The core convective velocity
follows the scaling of \citet{Christensen2010},
\begin{equation}
U_{\rm core} \sim \left(\frac{\ell\, q_{\rm conv}}{\rho\, H_t}\right)^{1/3}
\sim \left(\frac{g\,R_{\rm core}\,q_{\rm conv}}{\rho_{\rm core}}\right)^{1/3},
\label{eq:ucore}
\end{equation}
with temperature scale height $H_t = c_p/(\alpha g)$ and characteristic length
$\ell = R_{\rm core}$. Taking $q_{\rm conv}\sim q_{\rm tot}\sim M^{0.46}$,
$R_{\rm core} = R_{{\rm core},\oplus}\,\mu^{0.27}$, and core density
$\rho_{\rm core}\sim M_{\rm core}/R_{\rm core}^{3}\sim \mu^{0.19}$, Eq.~(\ref{eq:ucore})
gives
\begin{equation}
U_{\rm core} \sim (M^{0.5}M^{0.27}M^{0.46}M^{-0.19})^{1/3}\sim M^{0.347}
\end{equation}
or
\begin{equation}
U_{\rm core} = U_{{\rm core},\oplus}\,\mu^{0.347}.
\end{equation}

\citet{roberts2000geodynamo} recommend $U_{{\rm core},\oplus} = 5\times10^{-4}~{\rm m\,s^{-1}}$.

\subsection{Astronautical model: the multistage rocket equation}
\label{sec:rocket}
The Tsiolkovsky equation \citep{Tsiolkovsky1903} gives the velocity increment a
stage delivers,
\begin{equation}
\Delta v = I_{\rm sp}\, g_0 \,\ln\!\left(\frac{m_0}{m_f}\right),
\label{eq:tsiolkovsky}
\end{equation}
with $g_0 = 9.80665$~m\,s$^{-2}$. For a stage carrying mass $m_{\rm above}$ and
structural fraction $\epsilon$ (set to $0.10$) of the wet mass, the initial and
final masses are $m_0 = m_{\rm wet} + m_{\rm above}$ and
$m_f = \epsilon\, m_{\rm wet} + m_{\rm above}$. Defining
$E_t \equiv \exp(\Delta v_i/I_{\rm sp} g_0)$, Eq.~(\ref{eq:tsiolkovsky})
inverts to
\begin{equation}
m_{\rm wet} = m_{\rm above}\,\frac{E_t - 1}{1 - \epsilon E_t}.
\label{eq:mwet}
\end{equation}
For a fixed number of stages $n$ (see sec~\ref {sec:reliability}), the equation is solved iteratively, starting with the lightest stage, because $m_{\rm pay}$ is known; $m_{\rm above}$ is then solved sequentially once the upper stages are calculated. Equation~(\ref{eq:mwet}) also gives the
single-stage ceiling: as $m_{\rm wet}\to\infty$ (theoretical upper-limit), $\epsilon E_t \to 1$, so the
maximum attainable increment is $\Delta v_{\rm max} = I_{\rm sp} g_0\,
\ln(1/\epsilon)$. With $\epsilon = 0.10$ and $I_{\rm sp} = 450$~s (LH$_2$/LOX),
$\Delta v_{\rm max} \sim 10.2$~km\,s$^{-1}$, short of Earth's escape speed
($11.2$~km\,s$^{-1}$). On Earth, staging is mandatory for planetary escape. For $n>2$ stages we fix the top two stages at $I_{\rm sp} = 450$~s (LH$_2$/LOX) and the lower stages at $350$~s (RP-1/LOX) \citep[][Table 12.1.II]{HillPetersen1992}.

The total velocity budget is the sum of escape, gravity-loss, and drag terms
(the $\Delta v$ credit from planetary rotation is neglected, making the budget
conservative),
\begin{equation}
\Delta v = v_{\rm esc} + \Delta v_{\rm grav} + \Delta v_{\rm drag},
\qquad v_{\rm esc} = \sqrt{2GM/R},
\label{eq:dvbudget}
\end{equation}
with the gravity loss scaled to a conservative terrestrial value,
$\Delta v_{\rm grav} = 1500\,(g/g_\oplus)^{1/2}$~m\,s$^{-1}$ (the Space Shuttle
value is $1220$~m\,s$^{-1}$; \citealt[][Table 4-3]{SuttonBiblarz2001}). The
$g^{1/2}$ scaling follows from the constant-acceleration vertical ascent of
Sec.~\ref{sec:ascent}, where the net acceleration is $a_{\rm net}=0.5\,g$.
Gravity loss is $\Delta v_{\rm grav} \approx g\,t_{\rm burn}$. For a fixed
characteristic height $h$, the constant-acceleration kinematics from rest
give $h=\tfrac12\,a_{\rm net}\,t_{\rm burn}^{2}$, so with $a_{\rm net}=0.5\,g$
the burn time is
$t_{\rm burn}=\sqrt{2h/a_{\rm net}}=2\sqrt{h/g}$. Hence
$\Delta v_{\rm grav} \approx g\,t_{\rm burn}=2\sqrt{g\,h} \sim g^{1/2}$. The terrestrial
value sets the proportionality constant. Throughout the astronautical model we
take the surface gravity self-consistently with the adopted mass--radius
relation, $g = GM/R^{2} \approx g_\oplus\,\mu^{1-2\beta} = g_\oplus\,\mu^{0.46}$, i.e.\
the same gravity that sets $v_{\rm esc}$, gravity loss, and liftoff thrust. The
Valencia (2007) interior models use for the geophysical scalings
(Sec.~\ref{sec:geophys}) $g\propto\mu^{0.5}$. At the high-mass end
the two differ by $\sim\!10\%$, which shifts the engine-limit crossing by
$\lesssim\!0.5\,M_\oplus$ and leaves our conclusions unchanged.

\subsection{Atmospheric drag}
\label{sec:drag}
The drag force on the ascending stack is
\begin{equation}
F_d = \tfrac{1}{2}\,\rho_{\rm air}\, V^{2}\, S\, C_d,
\label{eq:drag}
\end{equation}
with $C_d = 0.2$ held fixed (a conservative choice; $C_d$ rises toward transonic
speeds, \citealt[][Fig.~10.4]{HillPetersen1992}). Air composition is fixed at
80\%~N$_2$/20\%~O$_2$ ($\mu_{\rm air} = 0.0288$~kg\,mol$^{-1}$) to limit
parameter count. Surface density follows the ideal gas law,
$\rho_{\rm surf} = P_{\rm surf}\,\mu_{\rm air}/(R\,T_{\rm surf})$ with
$T_{\rm surf} = 288$~K, and
\begin{equation}
\rho(h) = \rho_{\rm surf}\, e^{-h/H}, \qquad H = \frac{R\,T}{\mu_{\rm air}\, g},
\label{eq:expatmo}
\end{equation}
where $H$ is the scale height.

The frontal area $S$ is set by the first-stage propellant tank only: a cylinder
of RP-1/LOX with a mixture density of $\rho_f = 1000$~kg\,m$^{-3}$ and height-to-diameter
ratio of 3 \citep[][Table 12.1.I]{HillPetersen1992}. Engine bells, inter-tank
structure, and boosters are not modeled in $S$. The engines enter the propulsion
budget only through thrust and turbopump limits (Sec.~\ref{sec:enginesize}). They are treated as point masses and so do not contribute to cross-section calculations, though their required count is computed separately.

\subsection{Ascent integration}
\label{sec:ascent}
We discretize the atmosphere in steps $dh = 0.01\,H$ and integrate to $10\,H$
(density below $0.1\%$ of the surface value). \citet{HillPetersen1992} recommend a thrust ``from one and one-half to two times the initial weight in order that the vehicle may leave the ground with reasonable acceleration.'' We thus set $\mathrm{TWR} = 1.5$ as a balance between overcoming the thickest part of the atmosphere quickly but also imposing minimal stress on the structure. The net acceleration is thus $a_{\rm net} = 0.5\,g$ and, for vertical motion,
$V(h) = \sqrt{2\,a_{\rm net}\,h}$. First-stage wet mass $m_1$ from
Eq.~(\ref{eq:mwet}) and tank volume $V_{\rm tank} = m_1(1-\epsilon)/\rho_f$ yields
$d = (4 V_{\rm tank}/3\pi)^{1/3}$ and $S=\pi(d/2)^2$.

Propellant burn is simplified by considering a constant mean mass
$\langle m\rangle = m_0/2$, where $m_0$ is the \emph{gross} lift-off mass of the
full stack (all stages plus payload). The drag deceleration is then
\begin{equation}
a_d = \frac{F_d}{\langle m\rangle} = \frac{2 F_d}{m_0}.
\label{eq:adrag}
\end{equation}
Because $m_0$ and $m_1$ depend on $\Delta v_{\rm drag}$, we (crudely) seed $\Delta v_{\rm drag}$ as that of the Space Shuttle ($118\,P_{\rm surf}$~m\,s$^{-1}$)
\citep[][Table 4-3]{SuttonBiblarz2001}, optimize staging, recompute drag with
the updated $m_1$ and $m_0$, and repeat until $\Delta v_{\rm drag}$ changes by
less than 5\% (typically a few passes).

Our drag treatment is a first-order approximation rather than a coupled
trajectory solution. We prescribe a vertical ascent with constant net
acceleration, integrate $\Delta v_{\rm drag}$ along that velocity profile, and
iterate the result with the staging optimization once $m_0$ and $m_1$ are known
(Sec.~\ref{sec:reliability}). Over most of the
mass--pressure range explored here, drag contributes only a minor fraction of the
total ascent $\Delta v$ (Table~\ref{tab:results}; Sec.~\ref{sec:launchmass}), so
this simplification is unlikely to alter the principal conclusions, which are set
by escape speed, gravity losses, staging penalties, and engine-clustering limits.

\subsection{Engine sizing and clustering limit}
\label{sec:enginesize}
Engines must supply the propellant flow implied by liftoff thrust on the full
vehicle. There are two methods for pushing fuel from the fuel tanks into the combustion chamber. The first is to use pressurized gas, and the second is to use turbomachinery. Beyond a $\Delta v \sim 3$ km/s, turbomachinery becomes advantageous \citep[][p.~617]{HillPetersen1992}. The power of a turbopump can be computed from
\begin{equation}
P_{\rm turbo} = \frac{\dot{m}\,\Delta P_0}{\eta\,\rho_f},
\label{eq:turbo}
\end{equation}
with $\eta = 0.7$, $\rho_f = 1000$~kg\,m$^{-3}$, and $\Delta P_0 \sim 100$~bar
as the pump pressure rise for the RP-1/LOX engine (reproducing the published
F-1 turbopump power, $\sim\!45$~MW, to within $\sim\!18\%$ in our implementation;
\citealt{stangeland1992turbopumps}). Required thrust is
$F = \mathrm{TWR}\, m_0 g$ with $\mathrm{TWR}=1.5$, so
\begin{equation}
\dot{m}_{\rm tot} = \frac{F}{I_{\rm sp}\, g_0}
= \frac{1.5\, m_0\, g}{I_{\rm sp}\, g_0},
\label{eq:mdot}
\end{equation}
where $m_0$ is gross lift-off mass. We take $I_{\rm sp}=350$~s, matching the
lower-stage value in the staging model. While the actual F-1 sea-level $I_{\rm sp}$ is
less, using 350~s slightly \emph{under}states $\dot m_{\rm tot}$ and
hence $N_{\rm eng}$ (e.g.\ four versus five F-1 engines at Saturn-V scale).
We retain 350 s to be consistent with the stage model.
Each engine is considered to achieve the performance of the Rocketdyne F-1 ($\dot{m}_{\rm F\text{-}1} = 2600$~kg\,s$^{-1}$; \citealt[][p.~615]{HillPetersen1992}). With identical pump $\Delta P_0$ and $\eta$, fuel and oxidizer pumps per engine cancel in the ratio,
and
\begin{equation}
N_{\rm eng} = \left\lceil \frac{P_{\rm turbo,tot}}{P_{\rm turbo,eng}} \right\rceil
= \left\lceil \frac{\dot{m}_{\rm tot}}{\dot{m}_{\rm F\text{-}1}} \right\rceil .
\label{eq:neng}
\end{equation}
The $N_{\rm eng}\le 100$ limit is \emph{not} enforced during staging
optimization but on the reliability-optimal vehicle as a
feasibility constraint (Table~\ref{tab:assumptions}).

\subsection{Reliability-weighted stage selection}
\label{sec:reliability}

Equation~(\ref{eq:mwet}) implies that, in principle, any $\Delta v$ budget can be
met with enough stages: splitting the total increment reduces the per-stage
mass ratio and lowers $m_0$. To this end, we ask which $n$ a builder would choose when each stage adds complexity \emph{and} risk.

We model each stage as having an independent probability of success, $R_s$.
For an $n$-stage launch vehicle, the probability that all stages operate
successfully is therefore
\begin{equation}
R_{\mathrm{mission}}(n) = R_s^n.
\label{eq:rmission}
\end{equation}
Because a failed launch requires constructing and flying another vehicle, the
relevant quantity is not simply the mass of a single rocket, but the expected
mass expenditure required to achieve one successful mission. If a single mission
succeeds with probability $R_{\mathrm{mission}}$, then the expected number of
launch attempts required to obtain one success is
\begin{equation}
\frac{1}{R_{\mathrm{mission}}}.
\end{equation}
The expected mass expenditure per successful mission is therefore
\begin{equation}
M_{\mathrm{exp}}(n)
=
m_0(n)\,\frac{1}{R_{\mathrm{mission}}(n)}
=
m_0(n)\,R_s^{-n},
\label{eq:mexp}
\end{equation}
where $m_0(n)$ is the initial mass of an $n$-stage vehicle sized for the full
$\Delta v$ budget [Eq.~(\ref{eq:dvbudget})]. We therefore select the stage
count that minimizes the expected mass expenditure per successful mission:
\begin{equation}
n^\ast
=
\text{the value of } n \text{ that minimizes }
m_0(n)\,R_s^{-n}.
\label{eq:nstar}
\end{equation}
In practical terms, the optimal staging criterion continues to add stages only if each additional stage reduces the gross lift-off mass by more than the reliability penalty it introduces. In other words, adding another stage is advantageous only if
\begin{equation}
\frac{m_0(n+1)}{m_0(n)}
<
R_s.
\label{eq:rule}
\end{equation}
Equivalently, the fractional reduction in launch mass must exceed the per-stage
failure probability,
\begin{equation}
1 - R_s.
\end{equation}
For example, if the reliability of each stage is $R_s = 0.97$, then an
additional stage is justified only if it reduces the gross lift-off mass by more
than $3\%$.

Because the launch mass $m_0(n)$ decreases with increasing stage count, but does
so with diminishing returns, while the reliability penalty $R_s^{-n}$ grows
geometrically with increasing stage count, the expected mass expenditure [Eq.~(\ref{eq:mexp})] typically possesses a single well-defined minimum. The optimal stage count, therefore, occurs at the point where the marginal mass savings from adding another stage first become smaller than the associated reliability penalty.

This formulation replaces an arbitrary upper limit on stage number with a
physically motivated criterion determined by two measurable quantities: the
per-stage reliability and the incremental reduction in launch mass achieved by
staging. On more massive planets, larger per-stage $\Delta v_i$ keeps $m_0(n)$
falling steeply over more stages, so Eq.~(\ref{eq:rule}) is satisfied at larger
$n^\ast$ and $R_{\mathrm{mission}}(n^\ast)=R_s^{n^\ast}$ falls
(Table~\ref{tab:results}). We adopt $R_s=0.97$, representative of mature liquid
stages \citep{lawrence2017}; the qualitative trends depend weakly
on this choice.

\subsubsection{Implementation}

For each candidate $n$, we assign equal $\Delta v$ to every stage,
$\Delta v_i = \Delta v/n$, solve Eq.~(\ref{eq:mwet}) from the payload upward
(top stages first, using $I_{\rm sp}=450$~s for the upper two stages and
350~s below), and record $m_0(n)$. We evaluate $M_{\mathrm{exp}}(n)$ from
Eq.~(\ref{eq:mexp}) and increase $n$ until $M_{\mathrm{exp}}$ stops decreasing. After $n^\ast$ is fixed, $\Delta v_{\rm drag}$ is
recomputed with the corresponding $m_1$ and $m_0$ (Sec.~\ref{sec:ascent}) and
the search is repeated until drag converges.

\subsubsection{Limitations}

Reliability-optimal designs can still be impractically large. \emph{After}
choosing $n^\ast$ and reporting $m_0$, $N_{\rm eng}$, and
$R_{\mathrm{mission}}(n^\ast)$, we apply two filtering constraints from
Table~\ref{tab:assumptions}: $m_0 \le 4\times 10^5$~t (the ``Cheops pyramid'' limit
of \citealt{Hippke2018}) and $N_{\rm eng}\le 100$. A planet is
spacefaring-capable under our criteria only if the optimized vehicle passes
both tests.

\section{Model validation}
\label{sec:validation}

\begin{figure}[t]
\centering
\includegraphics[width=\linewidth]{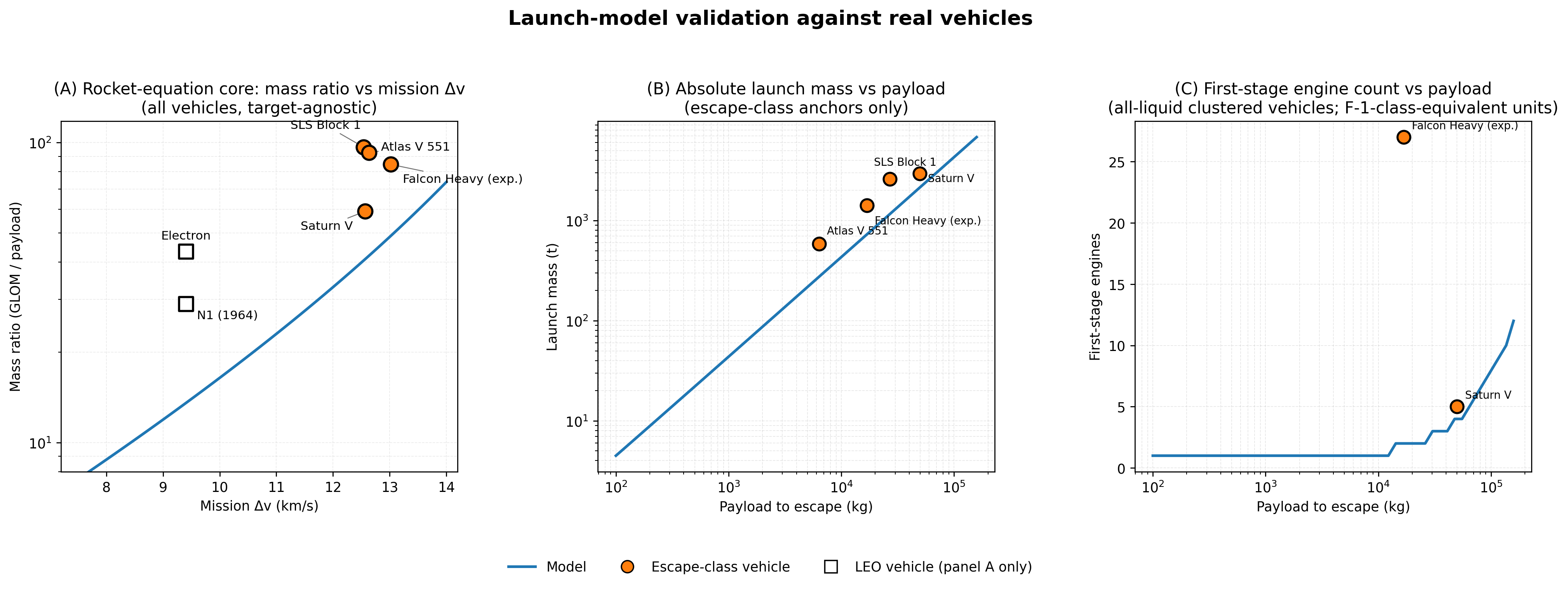}
\caption{Validation at $1\,M_\oplus$; vehicle data from
Table~\ref{tab:valrefs}.
(A) Mass ratio vs.\ mission $\Delta v$ for six real vehicles against the model
curve; (B) absolute launch mass for escape-class vehicles; (C) first-stage
engine count for all-liquid clustered vehicles.}
\label{fig:val}
\end{figure}

Before extrapolating to planetary masses other than Earth, we test whether the model
reproduces vehicles that have actually flown. We therefore evaluate it at
$1\,M_\oplus$, where the answer is known, and compare its predictions against six
vehicles spanning two orders of magnitude in size: the Saturn~V, Falcon~Heavy,
the Space Launch System (SLS) Block~1, the Atlas~V~551, the Soviet N1 (1964), and the
Electron small-launch vehicle (Fig.~\ref{fig:val}).
Vehicle masses, payloads, and stage or engine counts are taken from the sources
listed in Table~\ref{tab:valrefs} and mission $\Delta v$ budgets in
panel~A are computed as described below.
Saturn~V gross lift-off mass and spacecraft mass are means over Apollo~8--17
ground-ignition weights in \citet{Orloff2000}. Falcon Heavy gross lift-off mass and fully expendable payload to Mars are from an archived SpaceX specifications \citep{SpaceXFH}.
SLS~Block~1 gross lift-off mass and TLI payload are from the NASA SLS Reference Guide
\citep{NASASLS2022}. Atlas~V~551 gross lift-off mass are from \citet{AstronautixAtlas551} and escape payload ($C_3=0$; defined in Appendix~\ref{app:missionc3}) is from \citet{Schmidt2010}. N1 (1964) gross lift-off mass and LEO payload are from \citet{AstronautixN1}; here ``1964'' denotes the 1964 draft project, and because no N1 ever reached orbit its LEO payload is a design value. The flown 1969--1972 vehicle had a near-identical gross lift-off mass and the same 30-engine first stage. Electron gross lift-off mass and LEO payload are from \citet{RocketLabElectron}.

For panel~A we need a single propulsive budget from the launch site so that mass ratios can be compared on the same axis as the model's rocket-equation curve. We build that budget in two legs from the same loss model used elsewhere in this work (Secs.~\ref{sec:rocket}--\ref{sec:drag}). First,
ascent to a circular parking orbit at 200~km altitude requires
\begin{equation}
\Delta v_{\rm park} = v_{\rm circ}(r_{\rm park}) + \Delta v_{\rm grav}
  + \Delta v_{\rm drag},
\label{eq:dvpark}
\end{equation}
where $v_{\rm circ}=\sqrt{GM/r_{\rm park}}$, and $\Delta v_{\rm grav}$ and
$\Delta v_{\rm drag}$ follow Eqs.~(\ref{eq:dvbudget})--(\ref{eq:drag}) at
$1\,M_\oplus$ and 1~bar ($\Delta v_{\rm park}\approx 9.4$~km\,s$^{-1}$).
LEO-only vehicles (Electron, N1 (1964)) stop at $\Delta v_{\rm park}$.
For escape-class and interplanetary missions we add a prograde departure burn
from that parking orbit to the mission characteristic energy $C_3$,
\begin{equation}
\Delta v_{\rm dep} = \sqrt{\frac{2GM}{r_{\rm park}} + C_3} - v_{\rm circ}(r_{\rm park}),
\label{eq:dvdep}
\end{equation}
and take $\Delta v_{\rm mission}=\Delta v_{\rm park}+\Delta v_{\rm dep}$.
We set $C_3=0$ for Atlas~V~551 (Earth escape at $C_3=0$). For Falcon~Heavy
(Mars transfer) and SLS~Block~1 (lunar TLI), $C_3$ follows minimum-energy
Hohmann transfers for simplicity (Appendix~\ref{app:missionc3}). Saturn~V uses the mean
translunar cutoff $C_3$ from \citet{Orloff2000}. The validation scripts
reproduce the tabulated $\Delta v$ values from these rules.

\begin{table}[t]
\centering
\caption{Reference vehicle data for Fig.~\ref{fig:val} panel~A: gross lift-off
mass (GLOM), payload, effective mission $\Delta v$, and mass ratio $m_0/m_f$.
The model curve uses $\epsilon=0.10$, $I_{\rm sp}=350/450$~s, and $R_s=0.97$.}
\label{tab:valrefs}
{\tablefont\small
\begin{tabular}{@{}l l r r r r@{}}
\midrule
Vehicle & Mission & $\Delta v$ (km/s) & GLOM (t) & Payload (kg) & $m_0/m_f$ \\
\midrule
Saturn V & TLI & 12.6 & 2936 & 49\,800 & 59.0 \\
Falcon Heavy (exp.) & Mars transfer & 13.0 & 1421 & 16\,800 & 84.6 \\
SLS Block 1 & TLI & 12.5 & 2608 & 27\,000 & 96.6 \\
Atlas V 551 & Escape ($C_3=0$) & 12.6 & 587 & 6330 & 92.7 \\
N1 (1964) & LEO (design) & 9.4 & 2750 & 95\,000 & 28.9 \\
Electron & LEO & 9.4 & 13 & 300 & 43.3 \\
\midrule
\end{tabular}}
\end{table}

We first compare the most basic propulsive quantity, the ratio of gross lift-off
mass to payload mass, as a function of the mission velocity 
(Fig.~\ref{fig:val}A). Using each vehicle's actual mission $\Delta v$, the model
curve tracks the real vehicles across more than an order of magnitude in mass
ratio. Our model is a physics-based lower bound, because the flown vehicles lie above it. This is expected because their structural fractions are larger and their
stage-averaged specific impulses lower than the idealized values we adopt. That
the data parallel the model across the full $\Delta v$ range, rather than
diverging from it, indicates that the exponential mass--$\Delta v$ scaling at the
heart of the framework is correctly captured.

We next test the absolute launch mass but restrict the comparison to the escape-class vehicles for which our escape-trajectory assumptions apply: the Saturn~V, Falcon~Heavy, SLS~Block~1, and Atlas~V~551 (Fig.~\ref{fig:val}B). The model, an idealized lower bound, underpredicts their gross lift-off masses, with model/actual $\approx 0.45$--$0.51$ for Falcon~Heavy, SLS, and Atlas~V, and $\approx 0.73$ for the Saturn~V---its closest match. The first-stage engine-count comparison
(Fig.~\ref{fig:val}C) is restricted further to the all-liquid, engine-clustered
vehicles for which the model's F-1-class counting is meaningful (Saturn~V,
Falcon~Heavy); SLS and Atlas~V are excluded there because their liftoff thrust is
carried largely by solid boosters. For the Saturn~V the model reproduces the
first-stage engine count to within one engine when $I_{\rm sp}=350$~s is used
at liftoff (Sec.~\ref{sec:enginesize}). Agreement at $1\,M_\oplus$ indicates a
calibrated rather than unconstrained model. The turbopump power law of
Eq.~(\ref{eq:turbo}) is checked separately against the published F-1 turbopump
($\sim\!45$~MW; \citet{stangeland1992turbopumps}) in Sec.~\ref{sec:enginesize}. Taken together, these mass ratio, absolute launch mass, and engine-count comparisons indicate that the
model broadly reproduces flown hardware at $1\,M_\oplus$, and gives us license to extrapolate to super-Earths.

We next ask whether the model recovers the specific numbers quoted by
\citet{Hippke2018}, not only the qualitative trend (Table~\ref{tab:benchmark}).
Where Hippke gives a single-stage Tsiolkovsky estimate ($I_{\rm sp}=350$~s, no
structure), our model returns the same ratios ($26$ at $1\,M_\oplus$;
$\sim\!2700$ at his Kepler-20~b reference). When we turn on multistage
optimization without engineering caps, flown hardware lies above our idealized
lower bounds (Saturn~V: 68 vs.\ 41; Falcon Heavy: 83 vs.\ 49). At
$10\,M_\oplus$, Hippke's Apollo-class launch-mass estimate
($\sim\!4.0\times10^5$~t) agrees with ours ($\sim\!4.5\times10^5$~t) to within
$\sim\!13\%$. Imposing the engineering constraints refines Hippke's
order-of-magnitude chemical ceiling ($\lesssim\!10\,M_\oplus$) to a quantitative
$\sim\!11.5\,M_\oplus$ bound set by the 100-engine clustering limit.

\begin{table}[t]
\centering
\caption{Comparison of benchmarks from \citet{Hippke2018} against this model.
Hippke's $10\,M_\oplus$ cases use his Kepler-20~b scaling ($1.7\,R_\oplus$).
ESC = escape; launch masses in tonnes. ``Idealized'' runs our optimizer with
$\varepsilon=0.10$ and no reliability penalty or hardware caps; ``full model''
adds reliability-weighted staging, the 100-engine limit, and the
$4\times10^5$~t launch-mass budget.
$^*$: infeasible under our two engineering constraints.
The Saturn~V $m_0/m_f$ row compares our idealized model (41) against the
\emph{real} vehicle as quoted by Hippke (68); both are GLOM per unit injected
payload, the same quantity tabulated as $59$ (from Orloff) in
Table~\ref{tab:valrefs}. The model value is lower because it is a physics-based
lower bound, assuming lighter structure ($\varepsilon=0.10$) and higher
$I_{\rm sp}$ than the flown vehicle.}
\label{tab:benchmark}
{\tablefont\small
\begin{tabular}{@{}lrrr@{}}
\midrule
Benchmark & Hippke (2018) & idealized & full model \\
 & & (this paper) & (this paper) \\
\midrule
1-stage $m_0/m_f$, 1 $M_\oplus$ & 26 & 26 & --- \\
1-stage $m_0/m_f$, 10 $M_\oplus$ & 2700 & 2709 & --- \\
$v_{\rm esc}$, 10 $M_\oplus$ & 27.1 & 25.9 & 25.9 \\
Saturn~V $m_0/m_f$ (TLI) & 68 & 41 & 41 \\
Falcon Heavy $m_0/m_f$ & 83 & 49 & 49 \\
$m_0$ (t), 1 t ESC @ 1 $M_\oplus$ & --- & 44 & 44 \\
$m_0$ (t), 1 t ESC @ 10 $M_\oplus$ & --- & 10\,000 & 11\,000 \\
$m_0$ (t), 45 t ESC @ 10 $M_\oplus$ & 400\,000 & 453\,000$^*$ & 517\,000$^*$ \\
Max.\ $M_\oplus$, 1 t to ESC & $\lesssim 10$ & $\sim 25.7$ & $\sim 11.5$ \\
\midrule
\end{tabular}}
\end{table}

\FloatBarrier
\section{Results}
\label{sec:results}

\begin{table}[t]
\centering
\caption{Reliability-optimal vehicle to deliver a 1000~kg payload to escape.
Architecture columns (stages, engines, $R_{\rm mission}$) are at 1~bar surface
pressure (pressure has only a weak effect on launch mass for
$M_\oplus \gtrsim 4$; Fig.~\ref{fig:lm}); launch mass is listed at both 1 and
10~bar. The $f_{\rm drag}$ columns give the
drag contribution as a percentage of total required $\Delta v$
($v_{\rm esc}$ plus gravity and drag losses) at the indicated surface pressure.}
\label{tab:results}
{\tablefont\begin{tabular}{@{\extracolsep{\fill}}rrrrrrrrr}
\midrule
$M$ ($M_\oplus$) & $v_{\rm esc}$ (km/s) & stages & engines &
$R_{\rm mission}$ & \multicolumn{2}{c}{launch mass (t)} &
\multicolumn{2}{c}{$f_{\rm drag}$ (\% of total $\Delta v$)} \\
\cmidrule(lr){6-7}\cmidrule(lr){8-9}
& & & & & 1~bar & 10~bar & 1~bar & 10~bar \\
\midrule
0.5   &  8.7 &  2 &   1 & 0.94 &      17 &      22 &  0.9 &  8.1\\
1     & 11.2 &  2 &   1 & 0.94 &      44 &      53 &  0.4 &  3.9\\
2     & 14.4 &  4 &   1 & 0.89 &     170 &     183 &  0.1 &  1.4\\
4     & 18.6 &  6 &   3 & 0.83 &     809 &     834 &  0.0 &  0.5\\
6     & 21.5 &  8 &   9 & 0.78 &    2363 &    2403 &  0.0 &  0.2\\
10    & 25.9 & 11 &  55 & 0.72 &   11487 &   11571 &  0.0 &  0.1\\
11.5  & 27.3 & 11 &  98 & 0.72 &   19140 &   19256 &  0.0 &  0.1\\
\midrule
\end{tabular}}
\end{table}

\subsection{Launch mass: gravity dominates, pressure matters only on light worlds}
\label{sec:launchmass}

\begin{figure}[t]
\centering
\includegraphics[width=0.82\linewidth]{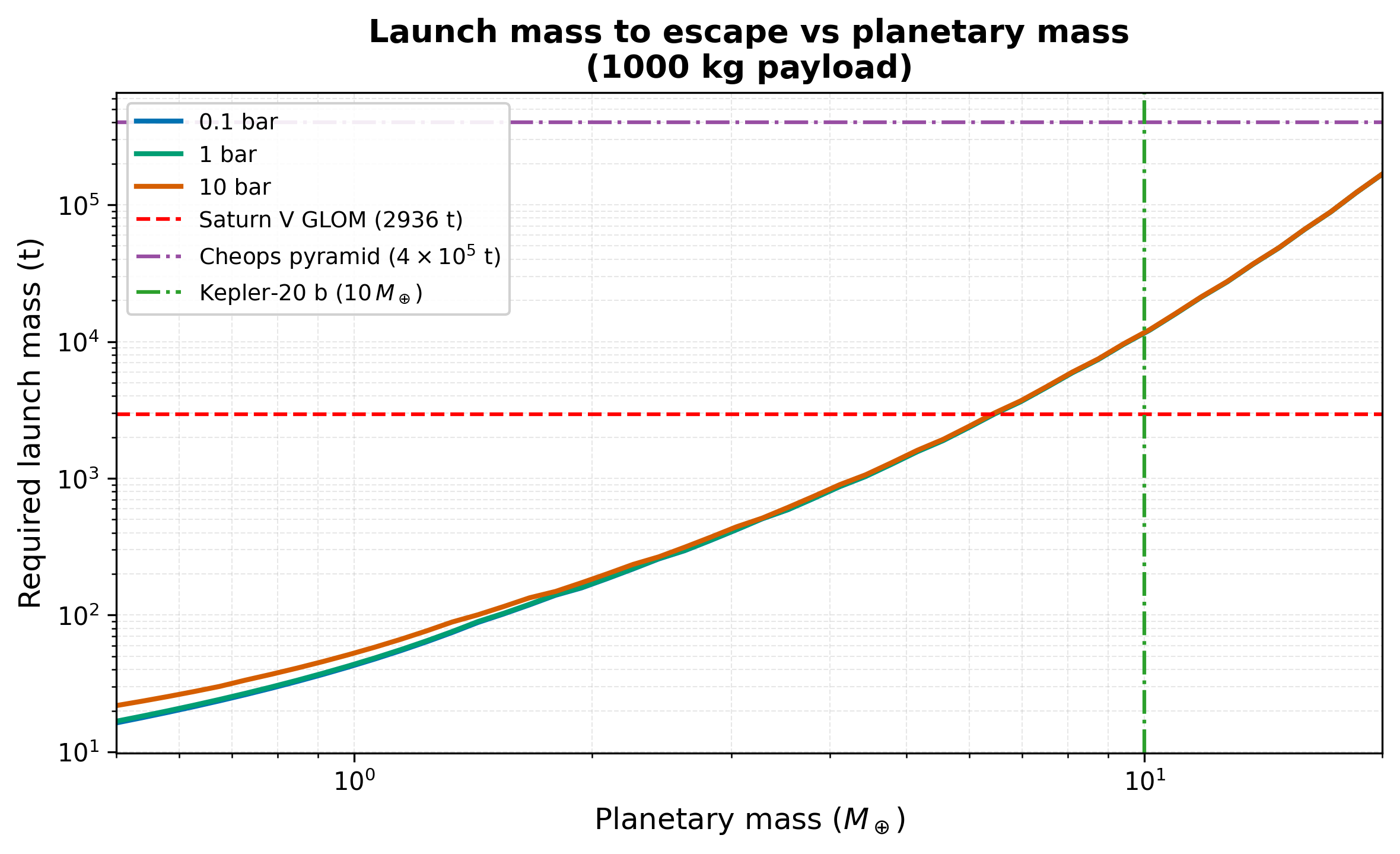}
\caption{Required gross launch mass (tonnes) to place a 1000~kg payload on an
escape trajectory, versus planetary mass, at $0.1$, $1$, and $10$~bar (blue,
green, orange). Log--log axes highlight the exponential growth with mass. Curves
separate at low $M_\oplus$ where drag is a larger fraction of the ascent
$\Delta v$; they merge for $M_\oplus \gtrsim 4$ where gravity losses dominate.
The red dashed line is the Saturn~V gross lift-off mass (2936~t, \citealt{Orloff2000}); the purple
dash--dot line is the Hippke ``Cheops pyramid'' launch-mass limit
($4\times10^5$~t). The green dash--dot vertical line marks Kepler-20~b
($10\,M_\oplus$; \citealt{Hippke2018}).}
\label{fig:lm}
\end{figure}

Figure~\ref{fig:lm} traces the gross launch mass required to deliver a
1000~kg payload to escape velocity, as a function of planetary mass, at three
representative surface pressures ($0.1$, $1$, and $10$~bar). Both axes use
logarithmic scaling so that the steep, rocket-equation growth with mass is
visible across the full $0.5$--$20\,M_\oplus$ range explored.

At Earth mass ($1\,M_\oplus$), all three curves lie near 44~t
(Table~\ref{tab:results}), consistent with the order of magnitude of historical
launchers to low Earth orbit and with the validation in Fig.~\ref{fig:val}.
Indeed, raising mass along the horizontal axis increases launch mass by orders of
magnitude: from $\sim\!17$~t at $0.5\,M_\oplus$ to $\sim\!2.2\times10^4$~t at
$12\,M_\oplus$ (1~bar). This vertical rise is the dominant visual feature of the
plot and reflects the combined growth of escape velocity, gravity losses, and
the reliability-weighted staging penalty.

The three pressure curves tell a second story. Notably, they \emph{diverge} at low
planetary mass and \emph{converge} at high mass. At $0.5\,M_\oplus$, moving from
$0.1$ to $10$~bar increases launch mass from $\sim\!16$~t to $\sim\!22$~t
($\sim\!35\%$; Table~\ref{tab:results}), because drag accounts for up to $\sim\!8\%$ of the total ascent
$\Delta v$ at 10~bar but only $\lesssim 0.2\%$ at 0.1~bar (Table~\ref{tab:results}).
By $4\,M_\oplus$ the same pressure swing changes launch mass by only
$\sim\!3\%$; by $6\,M_\oplus$ it falls to $\sim\!2\%$. Thus atmospheric
thickness is a first-order design variable only on low-gravity worlds, less so on
Earth, and essentially irrelevant on heavier super-Earths.

The dashed horizontal line marks the Apollo-era Saturn~V gross lift-off mass
(2936~t, mean Apollo 8--17, \citealt{Orloff2000}). The model crosses this benchmark near $\sim\!6\,M_\oplus$: a planet
heavier than this would require a launcher more massive than the Saturn V to place 1000 kg in an escape trajectory, even before the 100-engine and Cheops pyramid launch-mass constraints are applied. Together with Table~\ref{tab:results}, Fig.~\ref{fig:lm} shows that the
spacefaring envelope in launch-mass terms is set primarily by surface gravity,
with pressure acting as a modest modifier only where escape velocities are low.

\subsection{Engine count: the hardware ceiling}
\label{sec:engines}

\begin{figure}[t]
\centering
\includegraphics[width=0.82\linewidth]{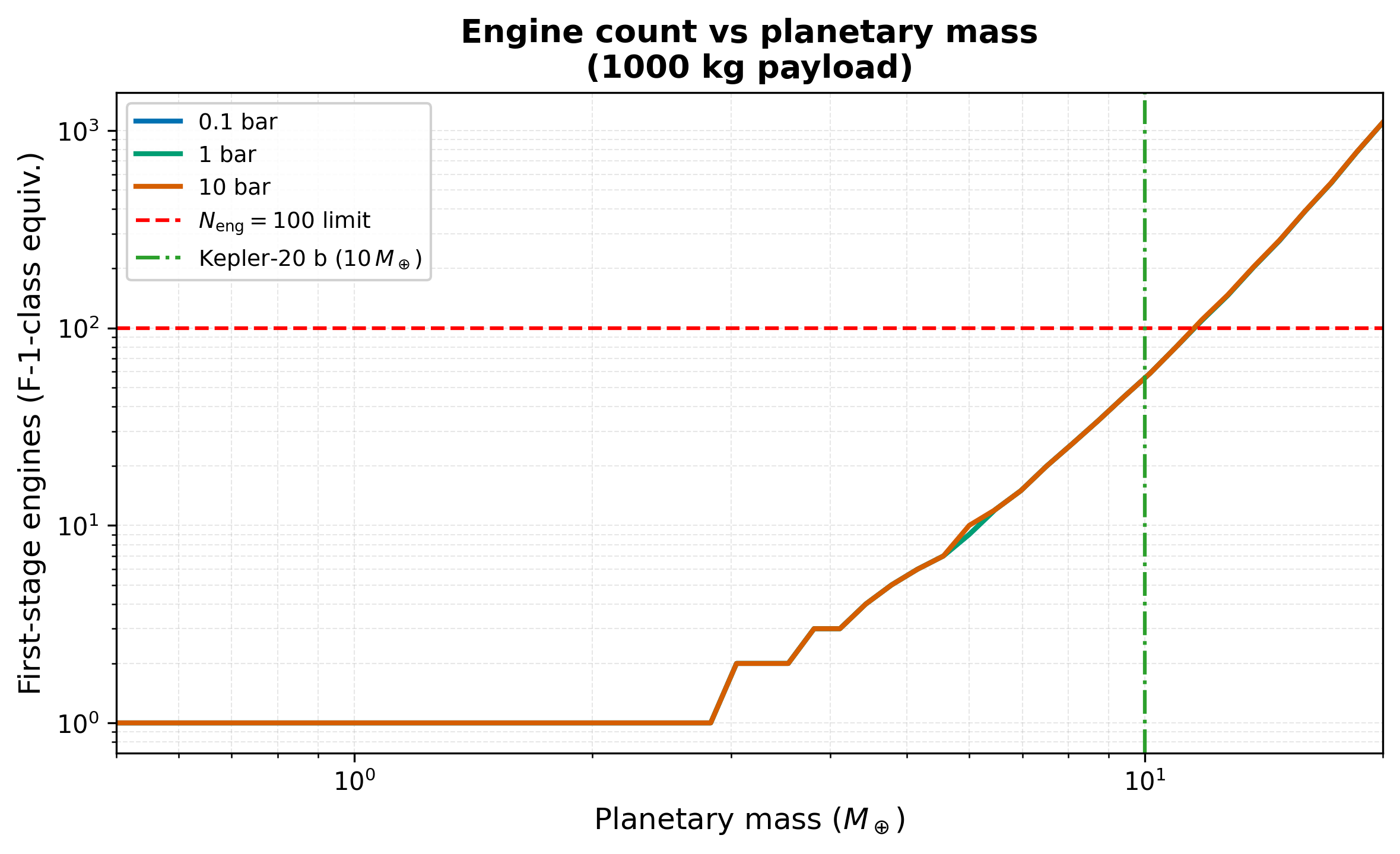}
\caption{First-stage engine count (F-1-class equivalents) for a 1000~kg escape
payload, versus planetary mass, at $0.1$, $1$, and $10$~bar. The three curves
overlap because pressure shifts the engine requirement only slightly at fixed
mass. The dashed line is the $N_{\rm eng}=100$ clustering limit, crossed near
$11.5\,M_\oplus$. The green dash--dot vertical line marks Kepler-20~b
($10\,M_\oplus$; \citealt{Hippke2018}).}
\label{fig:eng}
\end{figure}

Figure~\ref{fig:eng} shows the complementary quantity that ultimately bounds
chemical escape in our model: the estimated number of F-1-class engines on the
first stage, again plotted against planetary mass at $0.1$, $1$, and $10$~bar.
The same log--log format is used because engine count rises from order unity to
order $10^3$ across the mass range.

For $M_\oplus \lesssim 3$, a single engine suffices at all three pressures, because the
vehicle is small enough that one turbopump-limited thrust chamber can lift the
required mass (Table~\ref{tab:results}). Between $\sim\!3$ and $\sim\!11\,M_\oplus$
the count climbs steeply ($3$ engines at $4\,M_\oplus$, $9$ at $6\,M_\oplus$,
and $55$ at $10\,M_\oplus$), tracking the rapid growth of launch mass in
Fig.~\ref{fig:lm} and the increasing number of stages needed to limit
reliability-weighted propellant expenditure. Unlike launch mass, however, the
three pressure curves remain almost indistinguishable across the entire mass
range. At fixed $M_\oplus$, varying pressure from $0.1$ to $10$~bar changes the
engine estimate by at most a few engines. First-stage thrust is set mainly by
total vehicle mass and surface gravity, not by the modest differences in drag
that pressure introduces once $v_{\rm esc}$ is large.

The dashed horizontal line marks our adopted clustering limit,
$N_{\rm eng}=100$. The model crosses it near $\approx\!11.5\,M_\oplus$ (98 engines
at $11.5\,M_\oplus$ in Table~\ref{tab:results}), defining the practical upper
mass for escape of the 1000~kg benchmark under the hardware assumptions of
Sec.~\ref{sec:assumptions}. The 100-engine limit, anchored to the largest clusters flown ($27$--$33$ engines on Falcon~Heavy, the N1 (1964), and Super~Heavy),  is deliberately generous. Because $N_{\rm eng}$ rises steeply with mass, varying the limit from 50 to 200 engines shifts the crossing mass by only $\sim 2$--$3\,M_\oplus$. Notably, this ceiling reproduces the $\sim\!10\,M_\oplus$ fuel-ratio limit of \citet{Hippke2018} through an
independent, engine-counting argument rather than through the single-stage
Tsiolkovsky estimate alone. Mission reliability $R_{\rm mission}=R_s^n$ falls
from $0.94$ to $0.72$, quantifying how multistage risk compounds on more massive
worlds even before the engine limit is reached.

\section{Discussion}

Relative to \citet{Hippke2018} and \citet{Gonzalez2020}, this work adds three
elements. The reliability-weighted optimum [Eq.~(\ref{eq:nstar})] provides a criterion tied to measurable per-stage success and
launch mass. Figs.~\ref{fig:lm} and~\ref{fig:eng} show how launch mass and
$N_{\rm eng}$ scale with planetary mass at fixed pressure. Validation against
Saturn~V, Falcon~Heavy, and the F-1 turbopump is new for this
problem. The $\sim 11.5\,M_\oplus$ ceiling is consistent with Hippke, but the path
to it passes through testable engineering quantities rather than a single-stage
fuel-ratio estimate alone.

We included a geophysical assessment of plate tectonics and a dynamo (Sec.~\ref{sec:geophys}) so that spacefaring
habitability is not assessed on rocket physics alone. The non-dimensional
treatment, however, did not add additional constraints beyond those already imposed by the
coupled atmosphere--astronautics model over the masses we investigated
($0.5$--$20\,M_\oplus$). Scaling gives $Ra \propto (M/M_\oplus)^{1.71}$ and a
comparable growth of $Rm$ with mass, so both numbers remain above their adopted
onset thresholds for robust mantle convection and core dynamo action across
that range (Fig.~\ref{fig:geophys}). Mars ($\sim\!0.11\,M_\oplus$) likewise
passes both thresholds in the model yet lacks present-day plate tectonics and
an active dynamo. Mars thus serves as a reminder that our non-dimensional geophysical assessment is a first pass, not an evolution model. Therefore, plate tectonics and a magnetic field are labeled as plausible at every mass where we compute launch vehicles, and the spacefaring envelope is set by gravity, staging, drag, and engineering constraints
rather than by geophysical constraints. Models closer to geophysical reality could challenge this assertion.

We identify as spacefaring those rocky planets that plausibly sustain tectonics and a dynamo while permitting escape of a 1000~kg payload by chemical rocket. On planets outside this range, biology may persist without being spacefaring. The first-stage engine count exceeds our imposed $\sim 100$-engine clustering limit near $\sim 11.5\,M_\oplus$, over $0.1$--$10$~bar, which we treat as the practical upper bound for chemical escape of the benchmark payload. 

\section{Conclusion}

We developed a coupled geophysical--atmospheric--astronautical framework and used it to map a \emph{spacefaring envelope} onto planetary mass and surface pressure. We find that gravity, staging, and engine clustering, rather than atmospheric drag or geophysics, set this envelope. We also find that surface pressure is a second-order influence on launch mass and engine count for $M_\oplus \gtrsim 4$, but can matter substantially on low-mass worlds where drag is a larger fraction of the ascent $\Delta v$. Because $f_{\rm drag}$ remains small over most of the explored grid (Table~\ref{tab:results}), refinements to the ascent trajectory would not be expected to revise the super-Earth envelope identified here.

\appendix
\section{Validation mission characteristic energies}
\label{app:missionc3}

Panel~A of Fig.~\ref{fig:val} assigns a characteristic energy
$C_3\equiv v_\infty^{2}$ (the square of the hyperbolic excess speed, i.e.\ twice
the specific orbital energy, so that $C_3=0$ is marginal escape and $C_3>0$ a
hyperbolic departure) to each
escape-class vehicle before applying Eq.~(\ref{eq:dvdep}). Atlas~V~551 uses
$C_3=0$. Saturn~V uses the mean flown translunar cutoff from
\citet{Orloff2000}. For SLS~Block~1 (lunar TLI) and Falcon~Heavy (Mars transfer)
we adopt minimum-energy Hohmann transfers in the two-body approximation
\citep{SuttonBiblarz2001}. These budgets are trans-lunar or trans-Mars departure \emph{injection energies} from parking orbit, not end-to-end LEO-to-capture orbit totals.

We assume circular, coplanar orbits; parking at $r_{\rm park}=R_\oplus+200$~km;
lunar radius $r_{\rm Moon}=384{,}400$~km; and Mars at heliocentric radius
$r_{\rm Mars}=1.524$~AU ($1~\mathrm{AU}=1.496\times10^{11}$~m,
$\mu_\odot=1.327\times10^{20}$~m$^3$~s$^{-2}$). For the geocentric
Earth--Moon transfer (SLS~Block~1),
\begin{equation}
a = \tfrac{1}{2}(r_{\rm park}+r_{\rm Moon}),\quad
v_{\rm inj}=\sqrt{\mu_\oplus\!\left(\frac{2}{r_{\rm park}}-\frac{1}{a}\right)},\quad
C_3 = v_{\rm inj}^{2}-\frac{2\mu_\oplus}{r_{\rm park}},
\label{eq:c3moon}
\end{equation}
giving $C_3=-2.04\times10^{6}$~m$^2$~s$^{-2}$ and
$\Delta v_{\rm dep}=3.13$~km\,s$^{-1}$ via Eq.~(\ref{eq:dvdep}). For the
heliocentric Earth--Mars transfer (Falcon~Heavy),
\begin{equation}
a = \tfrac{1}{2}(r_\oplus+r_{\rm Mars}),\quad
v_{\rm dep}=\sqrt{\mu_\odot\!\left(\frac{2}{r_\oplus}-\frac{1}{a}\right)},\quad
v_\infty = v_{\rm dep}-v_\oplus,\quad
C_3 = v_\infty^{2},
\label{eq:c3mars}
\end{equation}
with $v_\oplus=\sqrt{\mu_\odot/r_\oplus}$, yielding
$C_3=8.68\times10^{6}$~m$^2$~s$^{-2}$ and
$\Delta v_{\rm dep}=3.61$~km\,s$^{-1}$. Combined with
$\Delta v_{\rm park}\approx 9.4$~km\,s$^{-1}$, the total mission budgets are
$12.5$~km\,s$^{-1}$ (SLS) and $13.0$~km\,s$^{-1}$ (Falcon~Heavy), as listed in
Table~\ref{tab:valrefs}.

\section*{Acknowledgments}
The author used AI-assisted software during the preparation of this manuscript:
the large language model Claude (Opus~4.8; Anthropic) was used intermittently in
May and June 2026. The study concept, scientific methods, and the underlying model were devised independently by the author. Working collaboratively with the author, the tool
was used to (i)~serve as a critical reader of the mathematical derivations; (ii)~develop the analysis code that produces the numerical results,
tables, and figures; (iii)~edit and revise the prose for clarity; (iv) assist in \LaTeX formatting. The author reviewed and verified all code and output, and
takes full responsibility for the content of this work.

\setlength{\bibsep}{0pt}%
\setlength{\bibparsep}{0pt}%
\setlength{\bibitemsep}{0pt}%

\end{document}